\documentclass[runningheads]{llncs}
\usepackage{cite}
\usepackage{amsmath}
\usepackage{amssymb,amsfonts}

\usepackage{mathtools, nccmath}
\usepackage{wrapfig}
\usepackage{comment}

\usepackage{graphicx}
\usepackage{comment}
\usepackage{textcomp}
\usepackage{xcolor}
\usepackage{caption}
\usepackage{subcaption}
\usepackage{float}
\usepackage{stfloats}
\usepackage{todonotes}
\usepackage{multirow}
\usepackage[thinlines]{easytable}
\usepackage{makecell}
\usepackage[normalem]{ulem} 
\usepackage{booktabs}
\usepackage{listings}
\usepackage{adjustbox}
\usepackage[obeyspaces]{url}
\usepackage{hyperref}

\usepackage{lstautogobble}

\definecolor{codegreen}{rgb}{0,0.6,0}
\definecolor{codegray}{rgb}{0.5,0.5,0.5}
\definecolor{codepurple}{rgb}{0.58,0,0.82}
\definecolor{backcolour}{rgb}{0.95,0.95,0.92}

\lstdefinestyle{mystyle}{
    backgroundcolor=\color{backcolour},   
    commentstyle=\color{codegreen},
    keywordstyle=\color{magenta},
    numberstyle=\tiny\color{codegray},
    stringstyle=\color{codepurple},
    basicstyle=\ttfamily\footnotesize,
    breakatwhitespace=false,         
    breaklines=true,                 
    captionpos=b,                    
    keepspaces=true,                 
    numbers=left, 
    frame=none,
    numbersep=5pt,                  
    showspaces=false,                
    showstringspaces=false,
    showtabs=false,                  
    tabsize=2
}

\DeclareMathOperator*{\argmin}{arg\,min}

\lstdefinestyle{Code_C_style}{
    commentstyle=\color{codegreen},
    keywordstyle=\color{magenta},
    numberstyle=\tiny\color{codegray},
    stringstyle=\color{codepurple},
    basicstyle=\ttfamily\footnotesize,
    breakatwhitespace=false,         
    breaklines=true,        
    frame=single,
    frameround=tttt,
    captionpos=b,                    
    keepspaces=true,                 
    numbers=left,                    
    numbersep=5pt,                  
    showspaces=false,                
    showstringspaces=false,
    showtabs=false,                  
    tabsize=2, 
    escapechar={|}
}
\usepackage{color}   \usepackage{soul}

\usepackage{ulem} 
\definecolor{darkgreen}{cmyk}{0.90,0,0.80,0.20}
\definecolor{darkred}{cmyk}{0, 1, 1, 0.10}
\definecolor{lightblue}{cmyk}{1, 0.70, 0, 0}

\definecolor{darkred}{cmyk}{0, 1, 1, 0.10}

\def\BibTeX{{\rm B\kern-.05em{\sc i\kern-.025em b}\kern-.08em
    T\kern-.1667em\lower.7ex\hbox{E}\kern-.125emX}}
    
\begin{document}

\title{Evaluation of Parameter-based Attacks against Embedded Neural Networks with Laser Injection}
\titlerunning{Published as a conference paper at SafeComp 2023}

\author{Mathieu Dumont\inst{1,2}\thanks{At SGS Brightsight after paper submission: mathieu.dumont@sgs.com.} \and Kevin Hector\inst{1,2} \and
Pierre-Alain Moëllic\inst{1,2}\thanks{Contact author.} \and
Jean-Max Dutertre\inst{3} \and 
Simon Pontié\inst{1,2}
}

\authorrunning{Published as a conference paper at SafeComp 2023}

\institute{CEA Tech, Centre CMP, Equipe Commune CEA Tech - Mines Saint-Etienne, F-13541 Gardanne, France \\
\email{\{name.surname\}@cea.fr}\\
\and Univ. Grenoble Alpes, CEA, Leti, F-38000 Grenoble, France 
\and Mines Saint-Etienne, CEA, Leti, Centre CMP, F-13541 Gardanne France
\email{dutertre@emse.fr}}



\maketitle

\begin{abstract}
Upcoming certification actions related to the security of machine learning (ML) based systems raise major evaluation challenges that are amplified by the large-scale deployment of models in many hardware platforms. Until recently, most of research works focused on API-based attacks that consider a ML model as a pure algorithmic abstraction. However, new implementation-based threats have been revealed, emphasizing the urgency to propose both practical \textit{and} simulation-based methods to properly evaluate the robustness of models. 
A major concern is parameter-based attacks (such as the Bit-Flip Attack – BFA) that highlight the lack of robustness of typical deep neural network models when confronted by accurate and optimal alterations of their internal parameters stored in memory. 
Setting in a security testing purpose, this work practically reports, for the first time, a successful variant of the BFA on a 32-bit Cortex-M microcontroller using laser fault injection. It is a standard fault injection means for security evaluation, that enables to inject spatially and temporally accurate faults. To avoid unrealistic brute-force strategies, we show how simulations help selecting the most sensitive set of bits from the parameters taking into account the laser fault model.
\end{abstract}

\keywords{Hardware Security, Fault Injection, Evaluation and certification, Machine Learning, Neural Network}

\setcounter{footnote}{0} 
\section{Introduction}
\label{introduction}
The massive deployment of Machine Learning (ML) models in a large spectrum of domains raises several security concerns related to their integrity, confidentiality and availability.
For now, most of the research efforts are essentially focused on models seen as abstractions, i.e. the attack surface is focused on so-called \textit{API-based attacks}, excluding threats related to their implementation in devices that may be physically accessible by an adversary, as it the case for embedded ML models.
The flaws intrinsically related to the models and the \textit{physical} ones may be used jointly to attack an embedded model or exploit data leakages.

A typical and well-known API-based attack are the \textit{adversarial examples}~\cite{szegedy2013}  that aims at fooling the prediction of a model with input-based alterations at inference time. 
However, 
recent works demonstrated that physical attacks are realist threats by targeting critical elements of a model such as activation functions~\cite{Hou2020} or its parameters (parameter-based attacks) \cite{rakin2019} as studied in this work. 

In such a security context, 
with the demonstration of worrying attack vectors, there is an urgent need of certification for AI-based systems more especially for \textit{critical} ones\footnote{See the European AI Act: \url{https://artificialintelligenceact.eu/}}.
Therefore, alongside the demonstration of new attacks and the development of defenses, an important challenge relies on the availability of proper robustness evaluation and accurate characterization methods in addition to both simulation and experimental tools and protocols. 
Model robustness evaluation is one of the most important challenge of modern artificial intelligence and several remarkable works from the Adversarial Machine Learning community have already raised major issues for adversarial examples~\cite{carlini2019evaluating}, with many defenses relying on weak evaluations~\cite{tramer2022detecting}.

Dealing with embedded neural network models and weight-based adversarial attacks, this challenge encompasses both \textit{safety} and \textit{security} concerns: parameters stored in memory may be altered by random faults because of hostile environments or strong energy consumption limitations~\cite{stutz2022random} and also be the target of an adversary that aims at optimally threatening a model.

\section{Related works and objectives}
\label{position}
\subsection{Parameter-based attacks}
\label{parameter_based_attacks}
\noindent\textbf{Implementation-based threats. }An important part of the ML security literature concerns \textit{algorithmic} or so-called \textit{API-based attacks} that exploit input/ouput pairs and additional knowledge from the model in case of white-box attacks. A large body of work shows that these threats concern every stage of the ML pipeline~\cite{papernot2018sok} and threaten the confidentiality (model and data), integrity and availability of the models. 
However, these attacks do not consider that the adversary may have a direct interaction with the algorithm as it can be the case for an embedded AI system. \textit{Implementation-based attacks} precisely exploit software or hardware flaws as well as the specific features of the device. For example, \textit{side-channel analysis}~\cite{DPAbook} have been demonstrated for model extraction as an efficient way to extract information from the model architecture or the values of the parameters~\cite{joud2023practical}. Alongside safety-related efforts that evaluate the robustness of ML models against random faults~\cite{stutz2022random}, some works demonstrate that models  are highly sensitive to \textit{fault injection analysis}~\cite{Barenghi2012} that alter the data, the parameters as well as the instructions flow \cite{Hou2020,DeepHammer2020}. 

\noindent\textbf{Weight-based adversarial attacks.} New attack vectors have been highlighted and more essentially parameter-based attacks (also named \textit{weight-based adversarial attacks}). 
Let's consider a supervised neural network model $M_W(x)$, with parameters $W$ (also referred as \textit{weights}), trained to optimally map an input space $\mathcal{X}=\mathbb{R}^d$ (e.g., images) to a set of labels $\mathcal{Y}$. $M$ is trained by minimizing a loss function $\mathcal{L}\big(M_W(x),y\big)$ (typically the cross entropy for classification task) that quantifies the error between the prediction $\hat{y} = M_W(x)$ and the correct label $y$. As formalized in~\cite{rakin2019} or~\cite{stutz2022random} with Eq.~\ref{eq_param_attack}, a parameter-based attack aims at maximazing the loss (i.e., increase mispredictions) on a small set of $N$ test inputs. As for the imperceptibility criterion of adversarial examples, the attacker may add a constrain over the perturbation by bounding the bit-level Hamming distance ($HD$) between the initial ($W$) and faulted parameters ($W'$), corresponding to an \textit{adversarial budget} $S$.  

\begin{equation}
    \underbrace{\max_{W'} \sum_{i=0}^{N-1}{\mathcal{L}\Big(M\big( x_i; W'\big),y_i\Big)}}_{\substack{\text{\sf \footnotesize \textcolor{black}{mispredictions}}}} \text{ s.t. } \overbrace{HD(W',W)\leq S}^{\substack{\text{\sf \footnotesize \textcolor{black}{adv budget}}}}
\label{eq_param_attack}
\end{equation}

A state-of-the-art parameter-based attack is the Bit-Flip Attack (hereafter BFA)~\cite{rakin2019} that aims to decrease the performance of a model by selecting the most sensitive bits of the stored parameters and progressively flip these bits until reaching an adversarial goal. In~\cite{rakin2019} or~\cite{he2020defending}, the objective is to ultimately degrade the model to a random-guess level. The selection of the bits is based on the ranking of the gradients of the loss w.r.t. to each bit $\nabla_b\mathcal{L}$, computed thanks to a small set of inputs. First, each selected bit is flipped (and restored) to measure the impact on the model accuracy. Then, the most sensitive bit is permanently flipped according to the gradient ascendant as defined in~\cite{rakin2019}.

\noindent\textbf{Adversarial goals. }Parameter-based attacks are not limited to the alteration of the target model integrity. BFA has been recently demonstrated for powerful model extraction in~\cite{rakin2022deepsteal} with an Intel i5 CPU platform: RowHammer is used to perform a BFA on the parameters of a model stored in DRAM (DDR3). The threat model in~\cite{rakin2022deepsteal} follows a typical model extraction setting: the adversary knows the model's architecture but not the internal parameters and has only access to a limited portion of the training dataset ($<10\%$). His goal is to build a \textit{substitute} model as close as possible as the target model. Interestingly, this joint use of RowHammer and BFA is performed in a side-channel analysis fashion: the observation of the induced faults enables to make assumptions on the value of some bits of the parameters. Then, the knowledge of these bits enables to efficiently train a substitute model, by constraining the value range of the parameters, with high \textit{fidelity} compared to the target model. We discuss this goal in Section \ref{reverse}.

\subsection{Scope and objectives} 
Because parameter-based attacks are the basis of both powerful integrity and confidentiality threats, their practical evaluation on the different platforms where fault injection may occur is becoming a critical need for present and future standardization and certification actions of critical AI-systems. We position our work on a different set of platforms than the main works related to the BFA (CPU platforms with DRAM), that is MCU platforms (Cortex-M with Flash memory), yet a very important family of embedded AI systems regarding the massive deployment of ML models on MCU-based devices for a large variety of domains. 
Our main positioning is as follows: 
\begin{itemize}
    \item We set this work in a security evaluation and characterization context. Therefore, we do not position ourselves through an \textit{adversary} but an \textit{evaluator} point of view.
    \item Our scope is parameter-based threats for neural network embedded in 32-bit microcontroller (hereafter MCUs).
    \item For that purpose, we use Laser Fault Injection (hereafter LFI) as an advanced and very spatially and temporally accurate injection means, a reference technique that is used in many security evaluation centers.
\end{itemize}

State-of-the-art is focused on simulation-based evaluations or on RowHammer attacks (i.e., exclusively DRAM platforms that excludes MCU). To the best of our knowledge, this work is the first to demonstrate the practicability and suitability of the characterization of a weight-based adversarial perturbation against Cortex-M MCU thanks to LFI.


\subsection{Related works}


Since the presentation of the BFA in~\cite{rakin2019}, several works analyzed the intrinsic mechanisms of the attack as well as potential protections~\cite{he2020defending, liu2022generating} and evaluated its properties according to the threat model, training parameters and model architecture~\cite{hector2022closer}. The standard BFA is \textit{untargeted} since the induced misclassifications are not chosen by the adversary. Therefore, some works also proposed \textit{targeted} versions~\cite{rakin2021tbfa} with specific target inputs and/or labels. Other recent works propose alternative methods to efficiently select the most sensitive parameters to attack \cite{stutz2022random}. For our work, we use and adapt (for LFI) the standard BFA of~\cite{rakin2019} as it is the state-of-the-art baseline for weight-based adversarial attacks. 

To the best of our knowledge, the only work related to laser injections for MCU-based platform against embedded neural networks has been proposed in~\cite{Hou2020}. 
Our work differs significantly by the target device and the elements we target in the model. In~\cite{Hou2020}, the authors used an 8-bit microcontroller (\texttt{ATMega328P}) and a 32-bit neural network implemented in C (a mulitlayer perceptron trained on the MNIST data set).
They only focused on the activation functions by inducing instruction skips (i.e., the faulted instructions are not executed, as if they were skipped). They used a laser and only targeted the last hidden layer.
We used a Cortex-M 32-bit microcontroller and embedded 8-bit quantized neural network thanks to a state-of-the-art open source library (\textit{Neural Network on Microcontrollers}\footnote{\url{https://github.com/majianjia/nnom/}}, hereafter NNoM) for model deployment. Our attack vector (BFA-like attack) and fault model enable to evaluate the robustness of a model against an advanced adversary that aims at significantly altering the performance of a model with a very limited number of faults or extract information about parameter values for model extraction. 

Thus, our work is also closely linked to~\cite{DeepHammer2020} that demonstrated at USENIX'20 a complete exploitation of the BFA with RowHammer on an Intel i7-3770 CPU platform. Although we target a different type of platforms with 32-bit MCU and another fault injection means (LFI), we share the same objective to go beyond simulations and propose a complete practical evaluation of a parameter-based attacks against embedded quantized DNN models.


\section{Evaluator assumptions and experimental setups} 
\label{security_testing}



\subsection{Goal and evaluator assumptions}
\label{threat_model}
\textbf{Objectives.} The main objective of the evaluator is to evaluate the robustness of a model against precise fault injections by decreasing the average accuracy on a labelled test set. More precisely, the scenario corresponds to a generic \textit{untargeted} case (i.e., the incorrect labels are not controlled by the evaluator).
Note that a targeted scenario (for specific test samples or target labels) is possible~\cite{rakin2021t} but out of the scope of our experiments. A secondary objective is to minimize the evaluation cost with a strategy that reduces the number of faults to be injected (i.e., avoid an exhaustive search that may be unrealistic according to the complexity of the target model). 

\noindent\textbf{Evaluator hypothesis.} Classically for security testing, the evaluator simulates a worst-case adversary that has a perfect knowledge of the model (white-box attack) and is able to query the model without limitation. The evaluator has a full access to the device (or clones of the device) and can perform elementary characterizations to adapt and optimize the fault injection set-up.

\subsection{Single bit-set fault model on Flash memory}
\label{bit-set_model}
We consider an accurate fault model relevant for LFI previously explained and demonstrated for NOR-Flash memory of Cortex-M MCU by Colombier and Menu~\cite{Colombier2019}: the bit-set fault model. As its name suggests, the fault sets a targeted bit to a logical \texttt{1}: when the bit was already at \texttt{1} the fault has no effect. 
When targeting a Flash memory at read time with a laser pulse, the induced bit-set is transient: it affects the data being read at that time while the stored value is left unmodified. Authors from~\cite{Colombier2019} explained the underlying mechanism of the bit-set fault injection with the creation of a \textit{photoelectric current} induced by the laser in a floating-gate (FG) transistor that flows from its drain to the ground. This current is added to the legitimate one so that the total current is above the reference that makes the bit read as a logical \texttt{1}.

LFI is a local fault injection means: its effect is restricted to the bit line connected to the FG transistor inside the laser spot area. More precisely several current components are induced in the affected transistors, depending on the laser spot diameter: up to two adjacent bits can be faulted simultaneously~\cite{Colombier2019}.

\subsection{Target and laser bench setup}
\label{laser_setup}

\textbf{Device under test (DUT).} Our target board embeds an ARM Cortex-M3 running at 8\,MHz. It includes 128\,kB of Flash memory and is manufactured in the $90$\,nm CMOS technology node. 
 The dimension of the chip is $3$ x $2.5$\,mm. Since LFI requires the surface of the die being visible, the microcontroller packaging was milled away with engraving tools to provide an access to its laser-sensitive parts. The chip was then mounted into a test board compatible with the ChipWhisperer CW308 platform.

\noindent\textbf{Laser platform.} Our laser fault analysis platform integrates two independent laser spots with a near infrared (IR) wavelength of $1,064$\,nm, focused through the same lens. Each laser spot has a diameter ranging from $1.5$ to $15 $\,µm depending on the lens magnification.
Both spots can move inside the whole field of view of the lens with minimum distortion.
The laser source can reach a maximum given power of $1,700$\,mW.
The delay between the trigger and the laser shot can be adjusted with a step of a few nanoseconds.
An infrared camera is used to observe the laser spot location on the target and a $XY$ stage enables to move the objective above the entire DUT surface.

\subsection{Datasets and models}
\label{NN_implem}

Although simulation-based works exclusively used complex deep neural networks trained for vision tasks (e.g.,  ResNet on ImageNet), it does not represent a large part of real-world applications that take benefit from classical fully-connected architecture (or multilayer perceptron, hereafter MLP) that fit and perform well on the widespread constrained platforms we studied in this paper. 

We considered two classical datasets. The \textbf{IRIS} dataset consists of 150 samples, each containing 4 real-value inputs and labelled according to 3 different iris species. We trained two simple models that provide an easy insight on the bit-set fault model effect: \texttt{IRIS\_A} and \texttt{IRIS\_B} are both composed with one hidden layer with one neuron and four neurons respectively. \texttt{IRIS\_B} has 96\% accuracy on the test set.    
\textbf{MNIST} dataset is composed of gray-scale handwritten digits images (28x28 pixels) from \texttt{0} to \texttt{9}. 
Our model (noted \texttt{MNIST}) is a MLP with one hidden layer of 10 neurons. Inputs are compressed to $\mathbb{R}^{50}$ with a classical principal component analysis. The resulting model has 620 trainable parameters and reaches 92\% of accuracy on the test set. All models use ReLU as activation function.

\subsection{Model implementation on MCU} 
\label{implem_nnom}

Models were trained with TensorFlow.
Few tools are available to embed previously trained models in microcontroller boards such as TensorFlow Lite, X-Cube-AI or NNoM. We chose NNoM as it is an efficient and convenient open source platform that fits our security testing objectives. NNoM offers 8-bit model quantization (with a standard uniform symmetric powers-of-two quantization scheme) and a complete white-box access to the inference code that enables to draw a timing profile of the sensitive operations we target (reading values from the Flash memory).
\vspace{-5pt}\vspace{-5pt}\vspace{-5pt}
\begin{figure}[h]
    \begin{subfigure}{0.48\textwidth}
    
    \begin{lstlisting}[language=C,  frame=single, basicstyle=\ttfamily\scriptsize, breaklines=true, autogobble]
while (rowCnt){
    //pA : address, stored input 
    //pB : address, stored weight
    for (int j = 0; j < dim_vec; j++){ //loop on all neuron parameters
        q7_t inA = *pA++;   //load input to inA, address increment
        q7_t inB = *pB++;  //load weight to inB, address increment
        ip_out += inA * inB; //neuron weighted sum
    }
    *pO++ = (q7_t)__NNOM_SSAT((ip_out >> out_shift), 8);
    rowCnt--;}  
\end{lstlisting}
    \caption{C code of the weighted-sum computation in a fully-connected layer.}
    \label{C_code}
    \end{subfigure}
    \hfill
    \begin{subfigure}{0.48\textwidth}
    \begin{lstlisting}[language={[x86masm]Assembler} ,frame=single, basicstyle=\ttfamily\scriptsize, breaklines=true, autogobble]
;q7_t inB = *pB++      ;Weight n+1 initialization
|\color{magenta}ldr|     r3, [r7, #80]  ;Loading the address of the weight n 
|\color{magenta}adds|    r2, r3, #1     ;Next weight address
str     r2, [r7, #80]  ;Input value loading into r2 reg
|\color{red}\textbf{ldrsb.w}| r3, [r3]       ;Weight value loading. LASER SHOT
|\color{magenta}strb|    r3, [r7,#23]   ;Store of the weight in SRAM reg
\end{lstlisting}
    \caption{Assembler code of line 6 of listing~\ref{C_code}. Our target is the load instruction, line 5.}
    \label{asm_code}
    \end{subfigure}
    \caption{C and Assembler codes from the NNoM inference.}
\end{figure}

\vspace{-5pt}\vspace{-5pt}\vspace{-5pt}
Listing~\ref{C_code} is an extract of the C code source of the core calculation of an inference from NNoM, that is the weighted sum between the inputs (i.e. the input data or the outputs of the previous layer) and the model parameters before the non-linear activation is applied. It consists in loading the neuron input and weight values (\texttt{inA} and \texttt{inB}, line 5 and 6), then process the multiplication and accumulation in an intermediate output value (\texttt{ip\_out}, line 7). Line 9 corresponds to the  quantization.
The assembler code in Listing~\ref{asm_code} corresponds to the weight initialization of Listing~\ref{C_code}, line 6. The weight value, stored into the Flash memory, is loaded into register \texttt{r3} (line 5 in red).
Based on our single bit-set fault model, if a laser beam is applied during the execution of the load instruction, a bit-set could be induced directly on the loaded value. 
To characterize the impact of the laser pulse, we synchronize the laser thanks to a trigger signal in the C code before line 6 and monitor the parameter value before and after the trigger using UART communication.

\section{A parameter-based attack with LFI}
\label{study}



To analyze the efficiency and practicality of LFI we first demonstrated the accuracy of the induced faults on a single neuron composed of four weights. Then, we scaled up to functional models trained on IRIS and MNIST to analyze the impact at a model-level. 

\subsection{Initial characterization on a neuron}
\label{bit-set_one_weight}

An important preliminary experience is to set up our laser bench on our DUT. For conciseness purpose, we do not detail all the Flash memory mapping procedure neither the selection of the laser parameters. For that purpose, we thoroughly followed the experimental protocol from~\cite{Colombier2019} and fixed the laser power to $170$\,mW and the pulse width to $200$\,ns. By selecting the lens magnification $\times5$, we chose a spot diameter of $15$\,µm to have a wide laser effective area.

\begin{figure*}[h!]
     \centering
     \begin{subfigure}[l]{0.46\textwidth}
         \centering
         \includegraphics[width=1\textwidth]{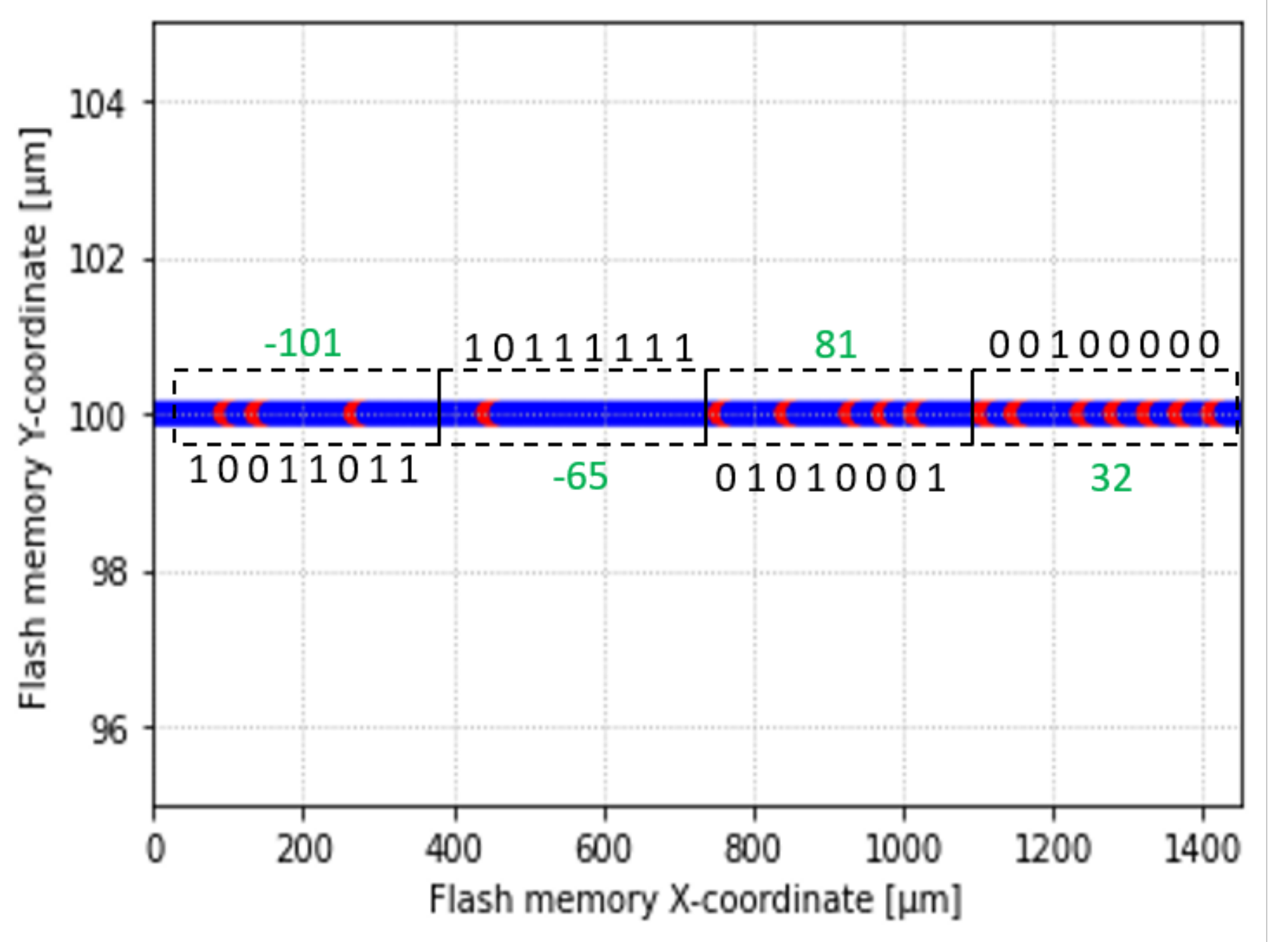}
	\caption{}
	\label{Multi_weight}
     \end{subfigure}
     \hfill
     \begin{subfigure}[r]{0.46\textwidth}
         \centering
         \includegraphics[width=1\textwidth]{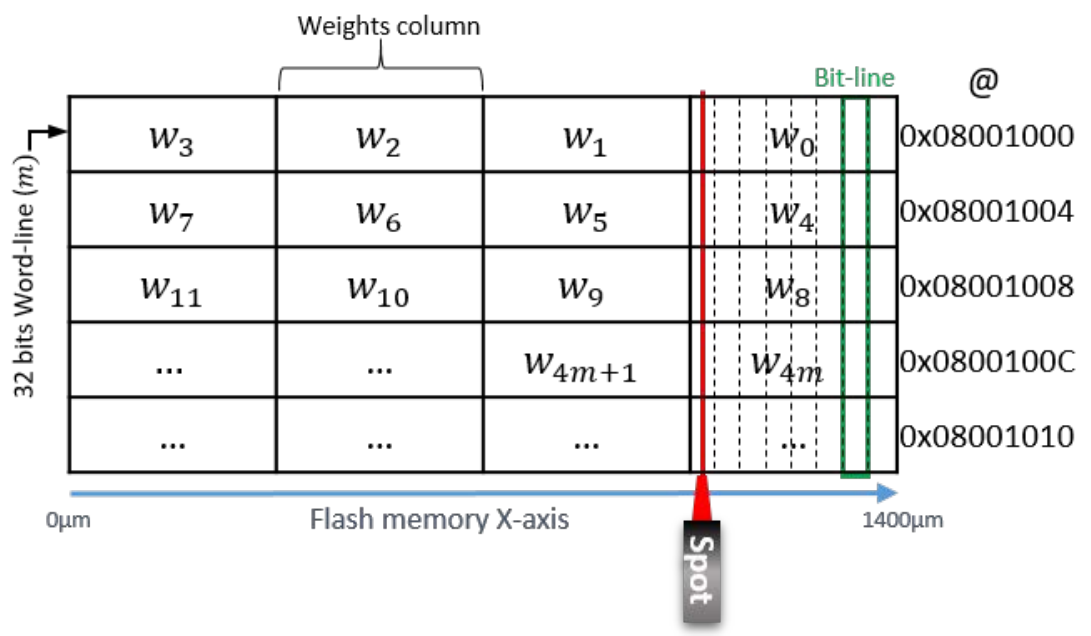}
	\caption{$m$ is the $m^{th}$ word-line}
	\label{flash_scheme}
     \end{subfigure}
     
     \caption{(a) Bit-set faults (red dots) induced on the 4 weights of a neuron (\texttt{IRIS\_A} model). (b) Flash memory schematic section filled with quantified weights.}
\end{figure*}

Then, we implemented a 4-weights neuron. We set the Y-position to $100$\,µm to only focus on the X-axis motion. Fig.~\ref{Multi_weight} shows that, when moving along the X-axis of the Flash memory, bit-sets are induced one after another on the whole 32-bit word line and the four quantized weights are precisely faulted. We repeated this experience with different weight values and noticed a perfect reproducibility, as also reported in \cite{Colombier2019}. 
By noticing positions of weights and their most significant bit (MSB), we easily conclude the little endian configuration of the Flash memory. Fig.~\ref{flash_scheme} illustrates how the weights are stored according to  the laser bench X-axis.
\vspace{-5pt}\vspace{-5pt}

\vspace{-5pt}\vspace{-5pt}

\subsection{Target multilayer perceptron models}
\label{one_model}

Since we can change the value of parameters related to a neuron, the next step consisted in analyzing the impact of such faults on the integrity of a target model and in measuring the potential drop of accuracy. For that purpose, we used the \texttt{IRIS\_B} model. For each X-step, we evaluated the robustness of the model by feeding it with 50 test samples (i.e., 50 inferences). Classically, we computed the adversarial accuracy by comparing the output predictions to the correct labels and measure the accuracy drop in comparison to the nominal performance without faults.  

During one inference, all the weight loading instructions triggered a laser shot (i.e. in total 40 shots for the hidden layer). By targeting one bit line at a given X-position of the laser, only weights on the same address column are faulted with a bit-set. For illustration, as pictured in Fig.~\ref{flash_scheme}, LFI actually induced bit-sets in the MSBs of weights $w_{0}$, $w_{4}$, $w_{8}$, $w_{4m}$ (with $m$ the $m^{th}$ word-line), which belong to the targeted bit line.

\begin{figure*}[h!]
     \centering
     \begin{subfigure}[l]{0.49\textwidth}
         \centering
         \includegraphics[width=1\textwidth]{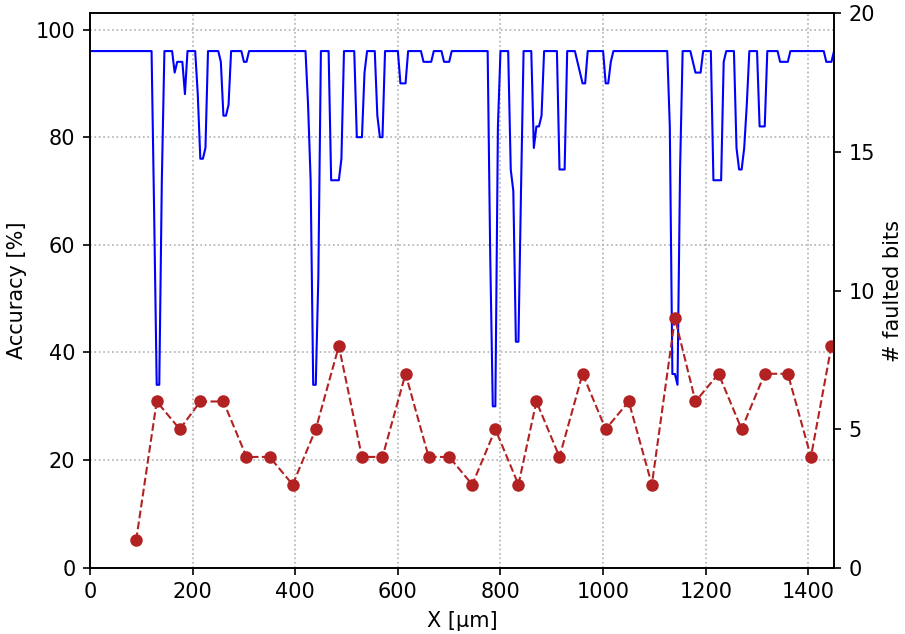}
	\caption{LFI with one spot}
	\label{Iris_charac_one_spot}
     \end{subfigure}
     \hfill
     \begin{subfigure}[r]{0.49\textwidth}
         \centering
         \includegraphics[width=1\textwidth]{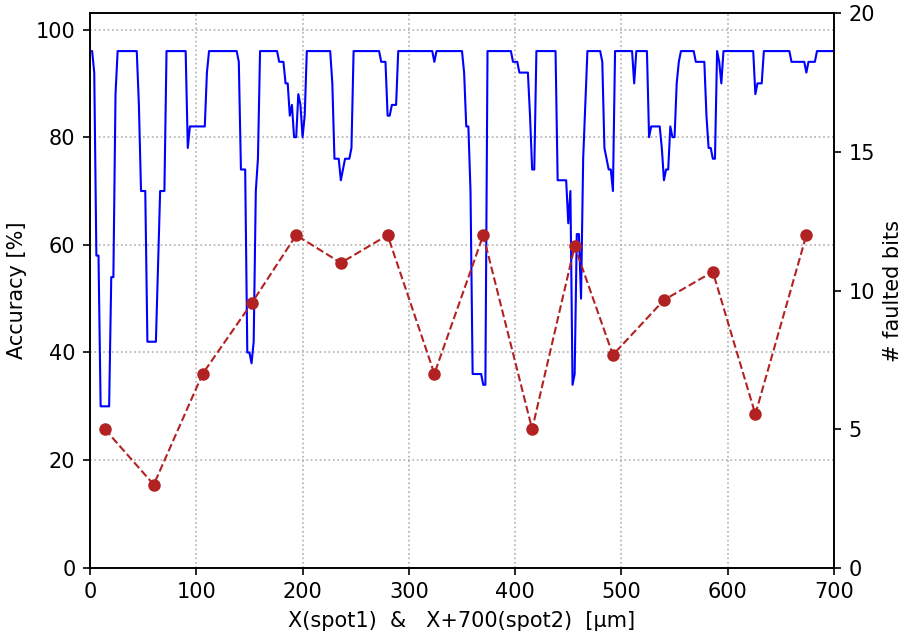}
	\caption{LFI with two spots}
	\label{bi_spot}
     \end{subfigure}
     
     \caption{Impact on the accuracy of LFI on \texttt{IRIS\_B} model. Average accuracy over 50 inferences (blue) and number of faulted bits (red).}
\end{figure*}

\vspace{-5pt}\vspace{-5pt}
Fig.~\ref{Iris_charac_one_spot} shows the impact of the laser shots on the model accuracy (blue curve) on the test set. The red curve provides the number of faulted bits. The X-axis corresponds to the position of the laser in the Flash memory.
Four regular patterns of accuracy drop corresponding to the four stored weights are clearly visible. The observed accuracy drops (5 to 6 per pattern) match the coordinates of the MSBs with a decreasing correlation.
For MSB bits, the model is close to a random-guess level (i.e. around 30\%). On average, around 6 bit-sets  are induced and the most harmful configuration corresponds to only 5 bit-sets (at $790$\,µm) leading to an adversarial accuracy of 30\%. 

One limitation is that only 1/4 of all weights could be faulted during one inference. However, since our laser platform integrates a second spot, we experimented a configuration where one spot targets the first bit line (i.e the MSB of the weight column $w_{3}$ in Fig.~\ref{flash_scheme}) and the second spot targets the bit line \#16 (i.e the MSB of the weight column $w_{1}$). As a result, two weight columns can be targeted during one inference, and half of the weights are likely to be faulted.
As reported in Fig.~\ref{bi_spot} more bit-set faults were injected (up to 12), and the accuracy drop for non-MSB bits is more important. 

\begin{figure}[t!]
    \centering
    \includegraphics[width=0.68\textwidth]{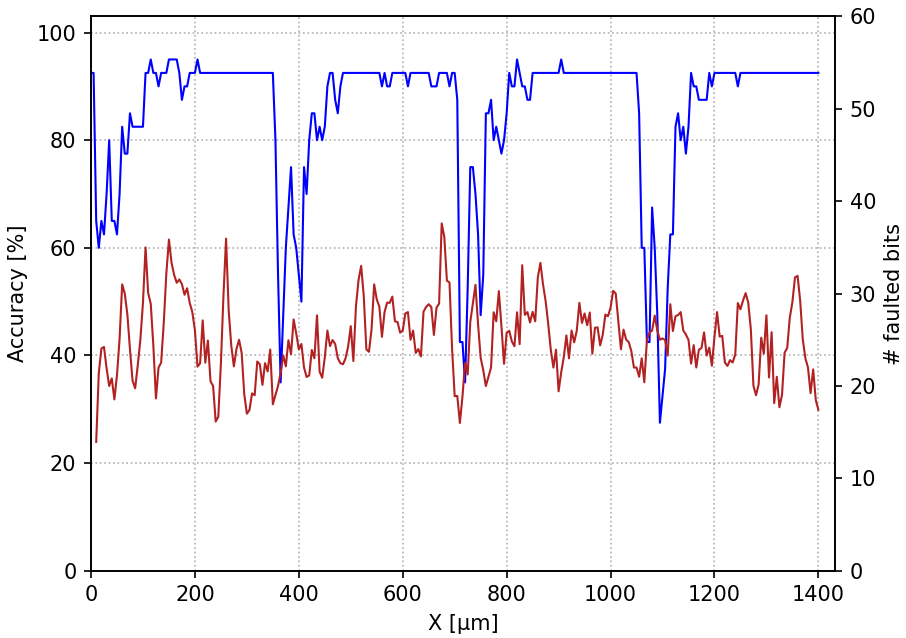}
    \caption{Impact on the accuracy of LFI on a MLP model trained on MNIST. Average accuracy over 100 inferences (blue) and number of faulted bits (red).}
    \label{mnist_charac}
\end{figure}


To extend the previous experiments, we embedded the \texttt{MNIST} model that gathers 620 weights (i.e. 4,960 bits) and strictly kept the same experimental setup.       
Results are presented in Fig.~\ref{mnist_charac}. 
On average, we observe more induced bit-sets with the most important drop of accuracy (65.5\%) reached with only 25 bit-sets  at $X= 1,100$\,µm. We also point out that, in some cases, a very slight improvement of the accuracy appeared. 

In comparison with \texttt{IRIS\_B}  (Fig.~\ref{Iris_charac_one_spot}), we note that bit lines are less distinctive. Indeed, even if the position of the most significant bits can be identified, other bits are difficult to distinguish.
Actually, in Fig.~\ref{flash_scheme} the bit line (green box) represents all single bit lines connected to the same bit index for different 32-bit words addresses. Since the \texttt{MNIST} model has more parameters, almost all bit lines are linked to a weight parameter value stored in Flash memory. The laser spot, with a size of $15 $\,µm, is larger than a single bit line, explaining why bit line indexes are hardly discernible.
Moreover, depending on the laser position, the effective area of the spot can encompass two different bit line indexes.

\subsection{Advanced guided-LFI}
\label{guided_lfi}
\textbf{Adapt BFA to LFI.}
So far, we exhaustively attacked all the parameters stored in memory. This \textit{brute-force} strategy may be impractical with deeper models.
Therefore, an important objective for an evaluator is to optimally select the most sensitive bits to fault.
For that purpose, as in~\cite{DeepHammer2020}, we adapted the bit selection principle of the standard BFA\footnote{\cite{rakin2019}: https://github.com/elliothe/BFA/} to our fault model.
We refer to this adapted attack as BSCA (\textit{Bit-Set Constrained Attack}). We also adapted the adversarial objective by introducing an \textit{adversarial budget}~\cite{hector2022closer, stutz2022random} representing a maximum of faults the evaluator is able to process. Indeed, the random-guess level objective of~\cite{rakin2019} overestimates the number of faults and raises variability issues because of the last necessary bit-flips needed to ultimately reach the objective~\cite{hector2022closer}. Therefore, we set the adversarial budget to 20 bit-sets.

As inputs, the BSCA has the target model $M_{W}$, its parameters $W$, the weight column index $m$ in the Flash memory, the bit line index $b$ and the adversarial budget $S$. The output of the BSCA is a new model $M_{W'}^{m,b}$ with $W'$ and $W$ that differ only by at most $S$ bit-sets. Then, we can compute an adversarial accuracy on $M_{W'}^{m,b}$ and measure the accuracy drop compared to the nominal performance of $M_W^{m,b}$. The BSCA proceeds through the following steps: 
\begin{enumerate}
    \item The BFA ranks the most sensitive bits of $W$ according to $\nabla_b\mathcal{L}$. 
    \item Exclude the bits already set to 1 and not related to $m$ and $b$.
    \item Pick the best bit-set and perform the fault permanently in $M$.
    \item Repeat the process until reaching $S$. The output is the faulted model $M_{W'}^{m,b}$.
\end{enumerate}
Finally, we perform BSCA over all the weight columns and bit line and keep the faulted model, simply referred as $M_{W'}$, with the worst accuracy ($Acc$) evaluated on a test set $\big(X^{test},Y^{test}\big)$: $M_{W'} = \argmin_{m,b} Acc( M_{W'}^{m,b}, X^{test}, Y^{test} )$

\begin{figure*}[t!]
     \centering
     \begin{subfigure}[l]{0.49\textwidth}
         \centering
         \includegraphics[width=1\textwidth]{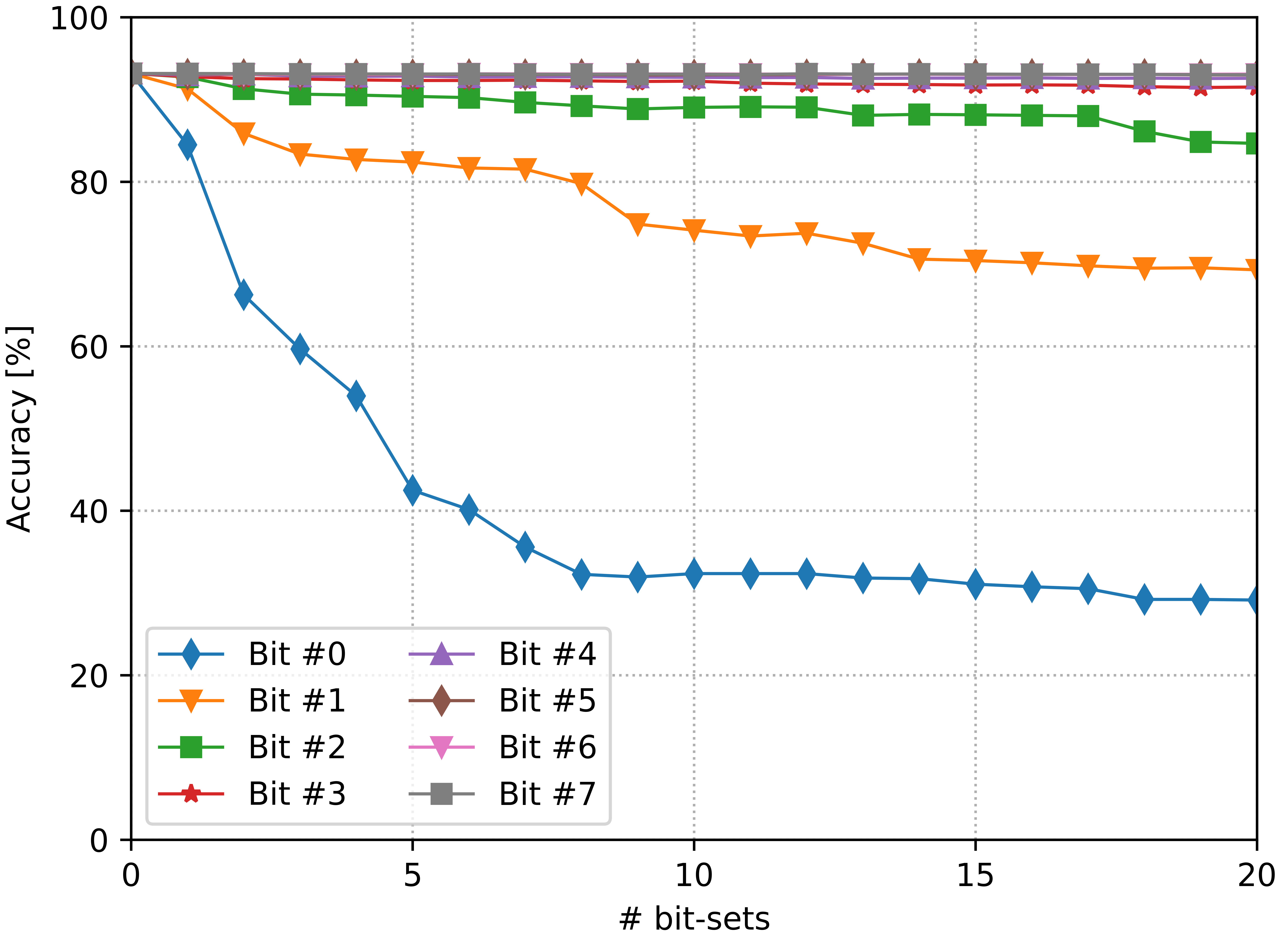}
	\caption{Influence of the faulty bit number on the accuracy for the weight column 2 (simulation results).}
	\label{simu_all_bits}
     \end{subfigure}
     \hfill
     \begin{subfigure}[r]{0.49\textwidth}
         \centering
         \includegraphics[width=1\textwidth]{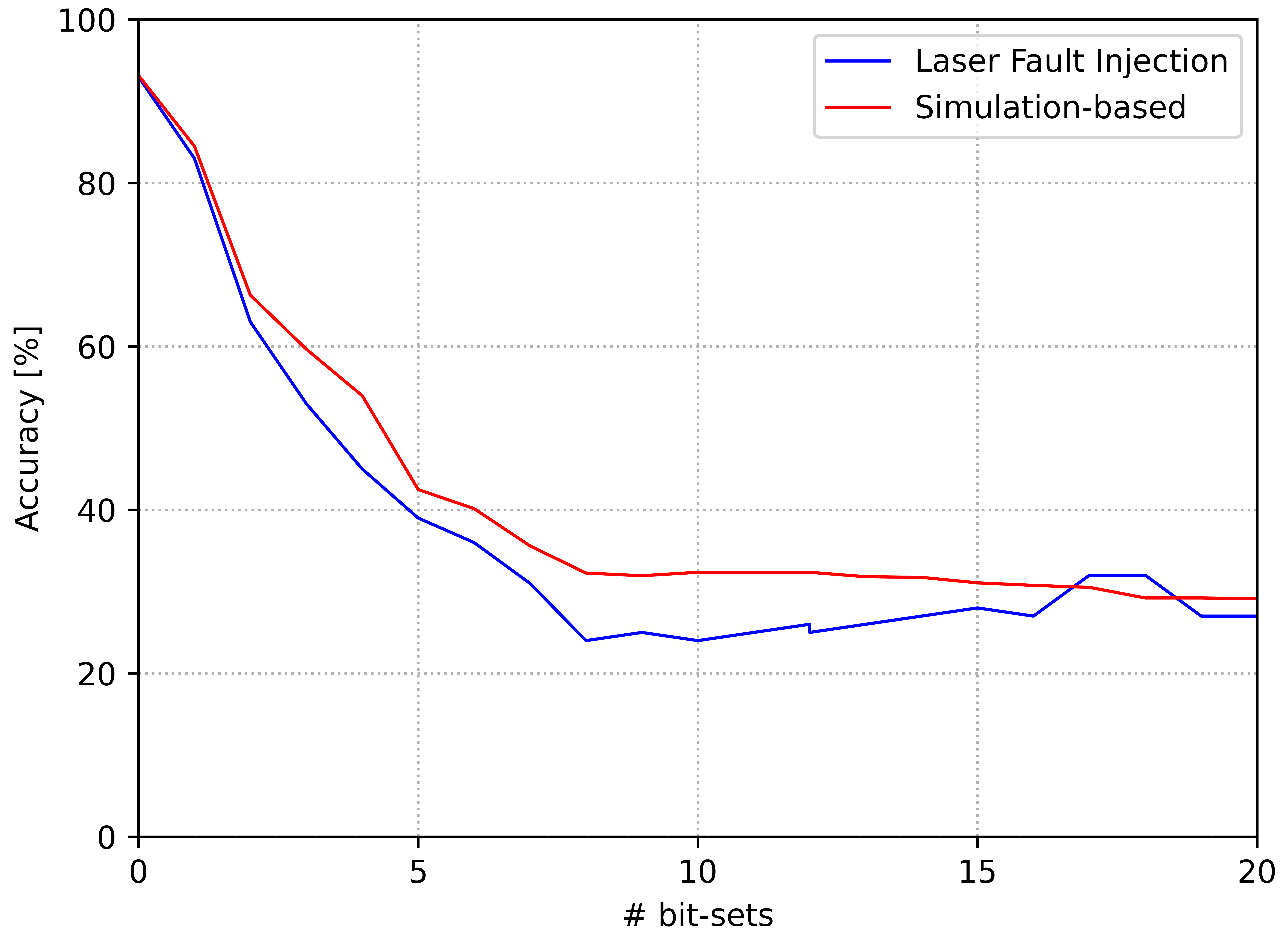}
	\caption{LFI BSCA attack targeting the 20 most sensitive MSB of the $2^{nd}$ weight column.}
	\label{simu_exp_simu}
     \end{subfigure}
     \caption{Guided-LFI on a MLP model trained on MNIST}
\end{figure*}

We used the \texttt{MNIST} model and simulated the BSCA to find the 20 most significant weights to fault for each weight column. For illustration purpose, Fig~\ref{simu_all_bits} shows the effect of bit-set induced on the 8 bits of a weight only for the second weight column (with the MSB refered as Bit~\#0). Among the most sensitive weights from the second column, only few bit-sets on the most significant bits efficiently alter the model performance, bit-sets on other bits have less or even no influence.

\noindent\textbf{Experiments and results. }The basic idea is to use the BSCA to guide the LFI. For that purpose, we need to put to the test that LFI reach (near) identical performance than what expected by simulations. We ran a BSCA simulation (in Python) over all the weight columns and bit lines that pointed out the MSB of the second column weight as the most sensitive.
Therefore, contrary to the previous experiments, the laser source was triggered only when the 20 most sensitive weights were read from the Flash.
The laser location was set accordingly ($X = 760$\,µm) and the power increased to $360$\,mW to ensure a higher success rate on weights stored in distant addresses.

The blue curve in Fig.\ref{simu_exp_simu} represents our experimental results (mean accuracy over 100 inferences) while the red one is the BSCA simulations for the MSB. First, we can notice that experimental and simulation results are almost similar, meaning that we can guide our LFI with high reliability and confidence\footnote{The fact that experimental results are slightly more powerful than simulations may be explained by the impact of the width of the laser spot on nearby memory cells.}.  
For an adversarial budget of only 5 bit-sets (0.1\% faulted bits) the embedded model accuracy drops to 39\% which represents a significant loss and a strong integrity impact compared to the nominal performance of 92\%.   
Moreover, after 10 bit-sets (accuracy to 25\%), the most effective faults have been injected and the accuracy did not decrease anymore. In a security evaluation context, this observation positions the level of robustness of the model according to the adversarial budget.
\vspace{-5pt}
\subsection{Exploitation of LFI for model extraction}
\label{reverse}
To illustrate the diversity of the adversarial goal an evaluator aims to evaluate (Section~\ref{parameter_based_attacks}), we propose a first insight of the use of LFI and BSCA for model extraction. At S\&P 2022, Rakin \textit{et al.}~\cite{rakin2022deepsteal}~demonstrated how to exploit BFA with RowHammer (i.e., exclusively DRAM platforms) in a model extraction scenario. The goal is to steal the performance of a model of which an adversary knows the architecture and a small part of the training data but not the internal parameters. The idea of~\cite{rakin2022deepsteal}~is to guess the value of the maximum of MSBs of the model thanks to rowhammer (bit-flip) and then to train a substitute model with these values as a training constraint.

Even though we rely on a different fault model (bit-set with LFI), it is relatively straightforward to follow the same extraction method as in~\cite{rakin2022deepsteal}. Indeed, by comparing the outputs of the model on a set of inputs (that could be random ones) with and without LFI we can guess the value of the targeted bit: if the \textit{faulted} outputs are not strictly identical than the nominal ones, that means the bit-set has an effect and therefore the bit was at \texttt{0}. Otherwise, the bit value was already to \texttt{1} or the bit-set has no influence on the outputs.
We simulated the BSCA to a deeper MNIST model with 3 layers with respectively 128, 64 and 10 neurons (109K parameters) and used 500 random input images to compare the outputs with and without bit-set faults on the MSB of each weight. To remove ambiguity, we made the simple assumption that working with enough inputs, a bit-set on a MSB at \texttt{0} has always an impact on the outputs. With that method, we managed to extract $91.9\%$ of the MSB of the model parameters, which is enough to efficiently trained a substitute model, consistently to \cite{rakin2022deepsteal}. Further analysis and experiments (more particularly, impact of the model architecture and the input set) should be the subject of future dedicated publications.

\section{Conclusion}
\label{discussions}


This work aims to contribute to the development of reliable evaluation protocols and tools for the robustness of embedded neural network models, a growing concern for future standardization and certification actions of critical AI-based systems.  
We conclude by highlighting some research tracks that may pursue this work and fill some limitations with further analysis on other model architectures and platforms (e.g., system-on-chip).

First, our results are not limited by the model complexity (BFA has been demonstrated on state-of-the-art deeper models with millions of parameters) but by the complexity of the MCUs. Indeed, the challenge is to characterize how the Flash memory is organized so that the evaluator can precisely target the weight columns and bit lines. Therefore, further experiments would be focused to other targets (e.g., Cortex M4 and M7). 

Second, contrary to adversarial examples with recent attacks specifically designed for robustness evaluation purposes~\cite{liu2022practical}, parameter-based attack still lacks of maturity and recent works highlight limitations of the BFA~\cite{stutz2022random, hector2022closer} and propose improvements or alternatives. Thus, considering or combining more attack methods would improve the evaluation by simulating a more powerful adversary.

Finally, this work could be widen to the practical evaluation of protections against weight-based adversarial attacks. Additionally to traditional generic countermeasures against fault injection~\cite{Barenghi2012}, specific defense schemes against BFA encompass weight clipping, clustering-based quantization\cite{Hou2020}, code-based detectors~\cite{hashtag2021} or adversarial training~\cite{stutz2022random}. To the best of our knowledge, none of these defenses have been practically evaluated against fault attack means such as RowHammer, glitching or LFI. As for adversarial examples (with so many defenses regularly \textit{broken} afterwards) the definition of proper and sound evaluations of defenses against parameter-based attacks is a research action of the highest importance.

\section*{Acknowledgment}
This work is supported by (CEA-Leti) the European project InSecTT\footnote{\url{www.insectt.eu}, ECSEL JU 876038} 
and by the French ANR in the \textit{Investissements d’avenir} program (ANR-10-AIRT-05, irtnanoelec);  and (MSE) by the ANR PICTURE program\footnote{\url{https://picture-anr.cea.fr}}. This work benefited from the French Jean Zay supercomputer with the AI dynamic access program.

\bibliography{bib/Biblio_MD,bib/Biblio_PA, bib/biblio}
\bibliographystyle{ieeetr}

\end{document}